\documentclass{JAC2003}
\usepackage{graphicx, subfigure,amsmath}
\usepackage{booktabs}

\begin{document}
\title{	Evaluation of the Beam Coupling Impedance of New Beam Screen Designs for the LHC Injection Kicker Magnets}
\author{H. Day\thanks{hugo.day@hep.manchester.ac.uk}$^{\dagger\ddagger\star}$, M.J. Barnes$^{\dagger}$, F. Caspers$^{\dagger}$, R.M. Jones$^{\ddagger \star}$, E. M\'{e}tral$^{\dagger}$, B. Salvant$^{\dagger}$ \\
$\dagger$ CERN, Switzerland,
$\ddagger$ School of Physics and Astronomy, The University of Manchester, Manchester, UK,\\
$\star$ Cockcroft Institute, Daresbury, UK \\}

\maketitle

\begin{abstract}
During the 2011 run of the LHC there was a significant measured temperature increase in the LHC Injection Kicker Magnets (MKI) during operation with 50ns bunch spacing. This was due to increased beam-induced heating of the magnet due to beam impedance. Due to concerns about future heating with the increased total intensity to nominal and ultimate luminosities a review of the impedance reduction techniques within the magnet was required. A number of new beam screen designs are proposed and their impedance evaluated. Heating estimates are also given with a particular attention paid to future intensity upgrades to ultimate parameters.
\end{abstract}


\section{INTRODUCTION}

The injection kicker magnets of the LHC (MKI) are transmission line kicker magnets used to inject beam from the transfer lines into the LHC. During the increase in beam current in the LHC in 2011 a significant increase in the temperature of the MKIs was observed \cite{mki-heating}. This was due to beam-induced heating as a result of the real component of the longitudinal beam coupling impedance. However, the MKIs have beam coupling impedance reduction measures in place to counteract this expected effect \cite{beam-screen}. 

An original beam screen for the MKI involved the use of an alumina "carrier" tube inserted into the MKI aperture, with 24 evenly distributed screen conductors inserted into open slots in the inner surface of the beam screen. These screen conductors are connected to the beam pipe at one end of the kicker magnet, and capacitively coupled to the beam pipe via an external metalisation on the ceramic tube at the other. The purpose of this capacitive coupling is to preserve the field properties of the kicker due to the strict requirements on field homogeneity and rise time ($\approx 1\mu{}s$).

\begin{figure}
\begin{center}
\includegraphics*[width = 0.35\textwidth]{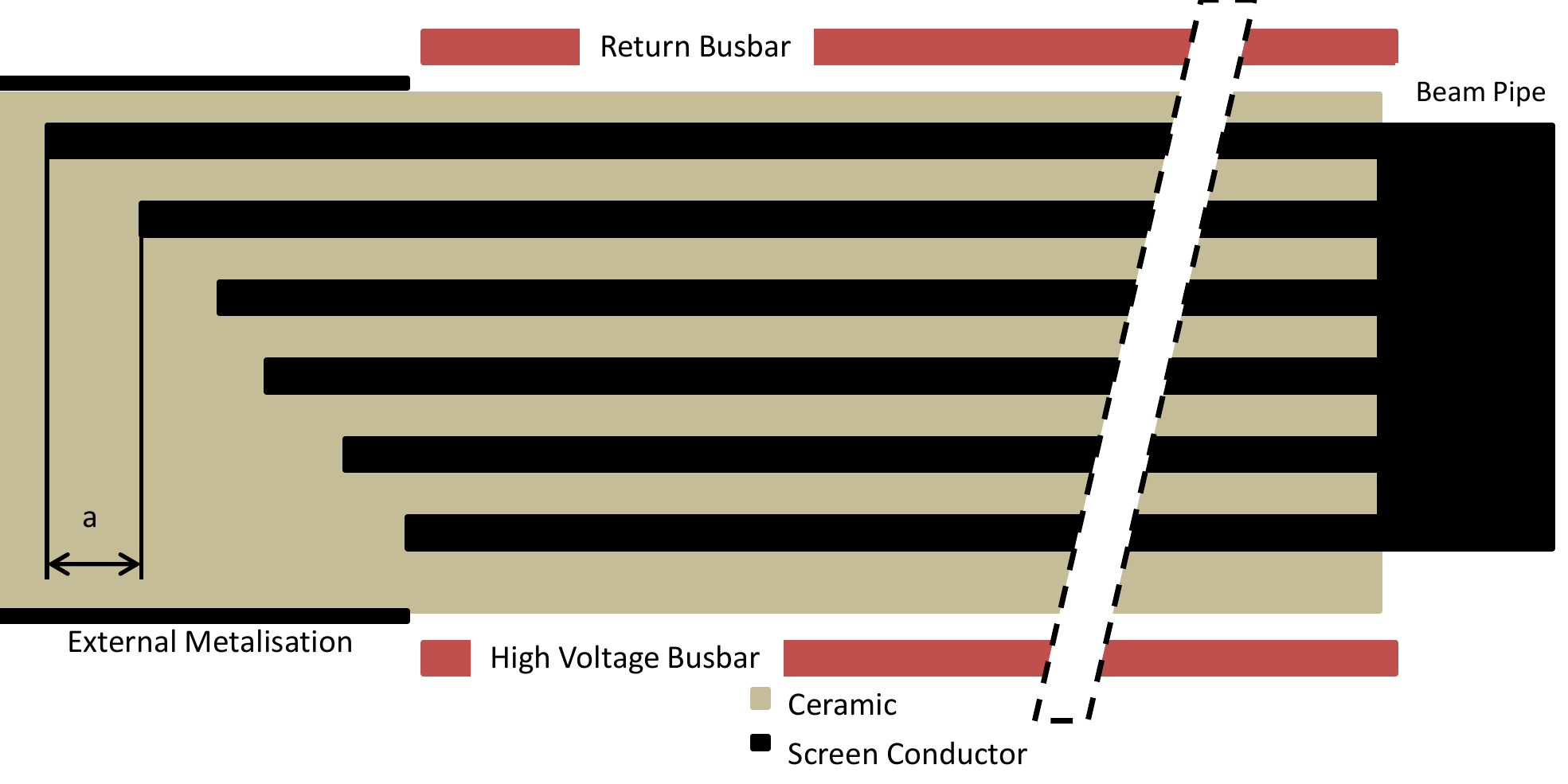}
\label{fig:short-screens}
\end{center}
\caption{Layout of the beam screen, with the screen conductors electrically connected to the beam pipe at one end of the screen, and capactively coupled to the beam pipe at the other by an overlapping external metallisation.}
\end{figure}

For most of the MKIs presently in use in the LHC, the 9 screen conductors closest to the HV conductor (see Fig.~1) have been removed due to issues of electric breakdown\cite{beam-screen}. This has meant that the beam screens have provided sub-optimal impedance reduction during normal operation of the LHC. In order to better understand the source of this heating a detailed campaign of measurements and simulations of the LHC-MKI have been carried out, with the intention of providing a number of different solutions to the present and any possible future heating problems.

\begin{figure}
\begin{center}
\subfigure[]{
\includegraphics*[width = 0.35\textwidth]{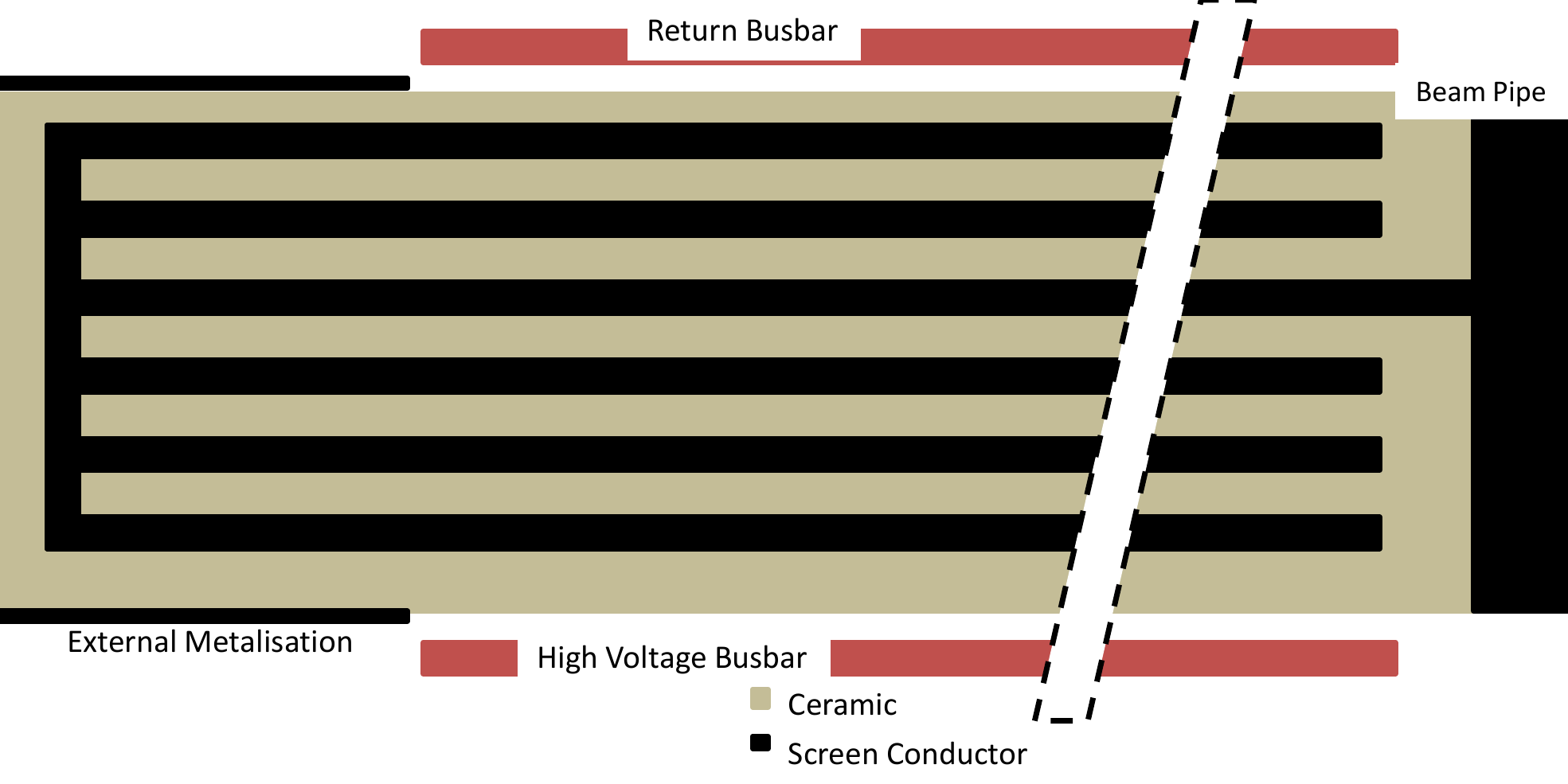}
\label{fig:alt-screen}
}
\subfigure[]{
\includegraphics*[width = 0.35\textwidth]{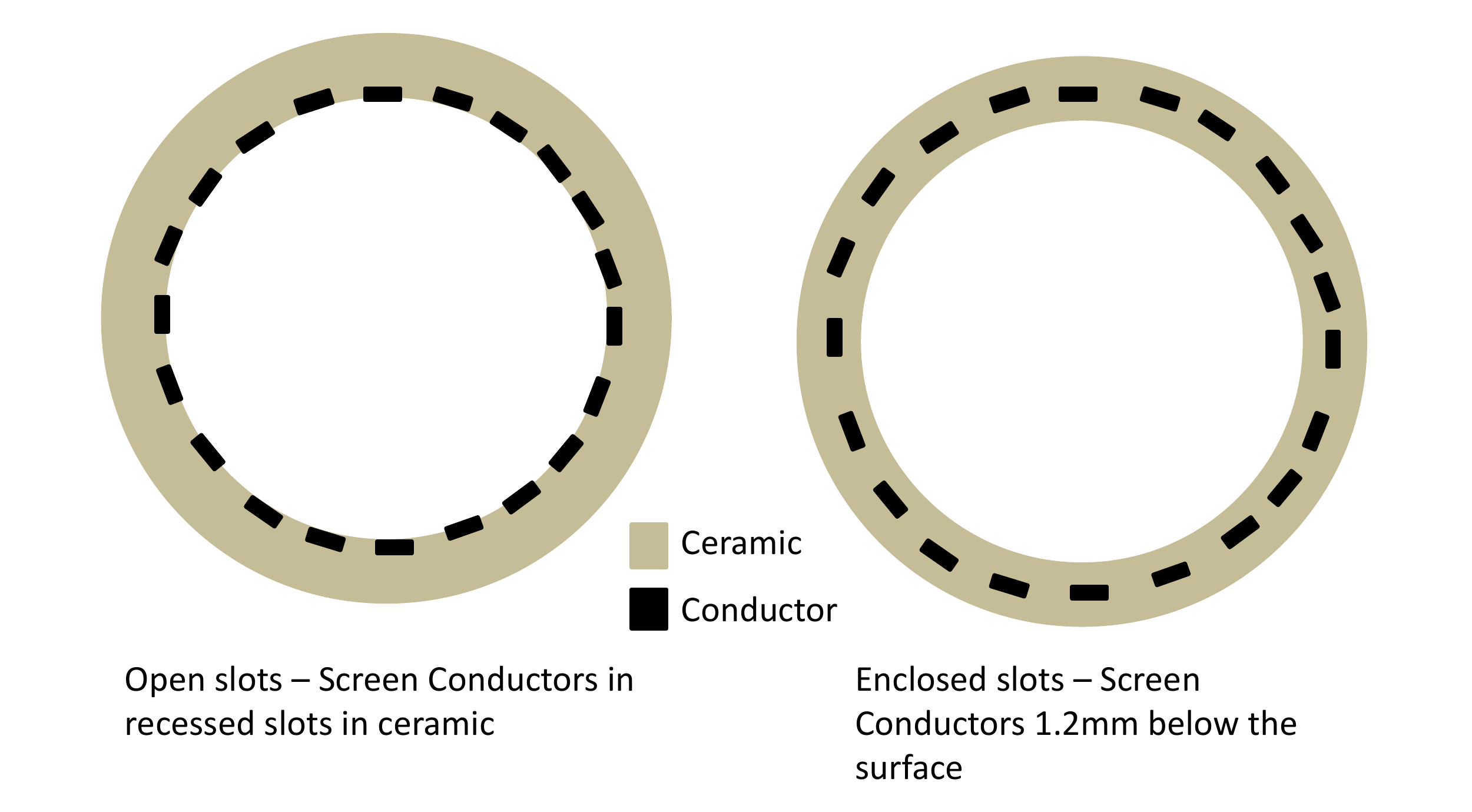}
\label{fig:enclosed_cond}
}
\end{center}
\caption{Cross sections of LHC-MKI beam screen with either \ref{fig:alt-screen} an alternative screen layout, in which most conductors are capactively coupled at both ends, and \ref{fig:enclosed_cond} the beam screen in which the screen conductors are in enclosed slots rather than open slots.}
\end{figure}

\section{BEAM COUPLING IMPEDANCE SIMULATIONS AND MEASUREMENTS}

\begin{figure}
\begin{center}
\includegraphics*[width = 0.4\textwidth]{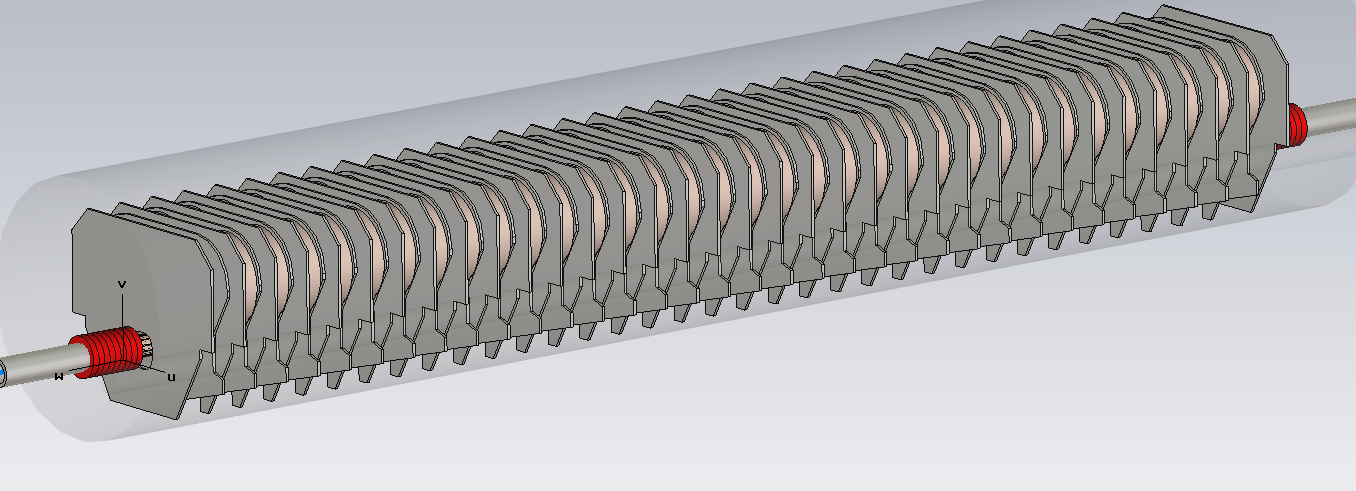}
\label{fig:mki_schematic}
\end{center}\caption{Schematic of the MKI. The tank is shown transparent, the red tiles are the damping ferrites, and the internal magnet structure can be seen.}
\end{figure}

\begin{figure}
\begin{center}
\includegraphics*[width = 0.4\textwidth]{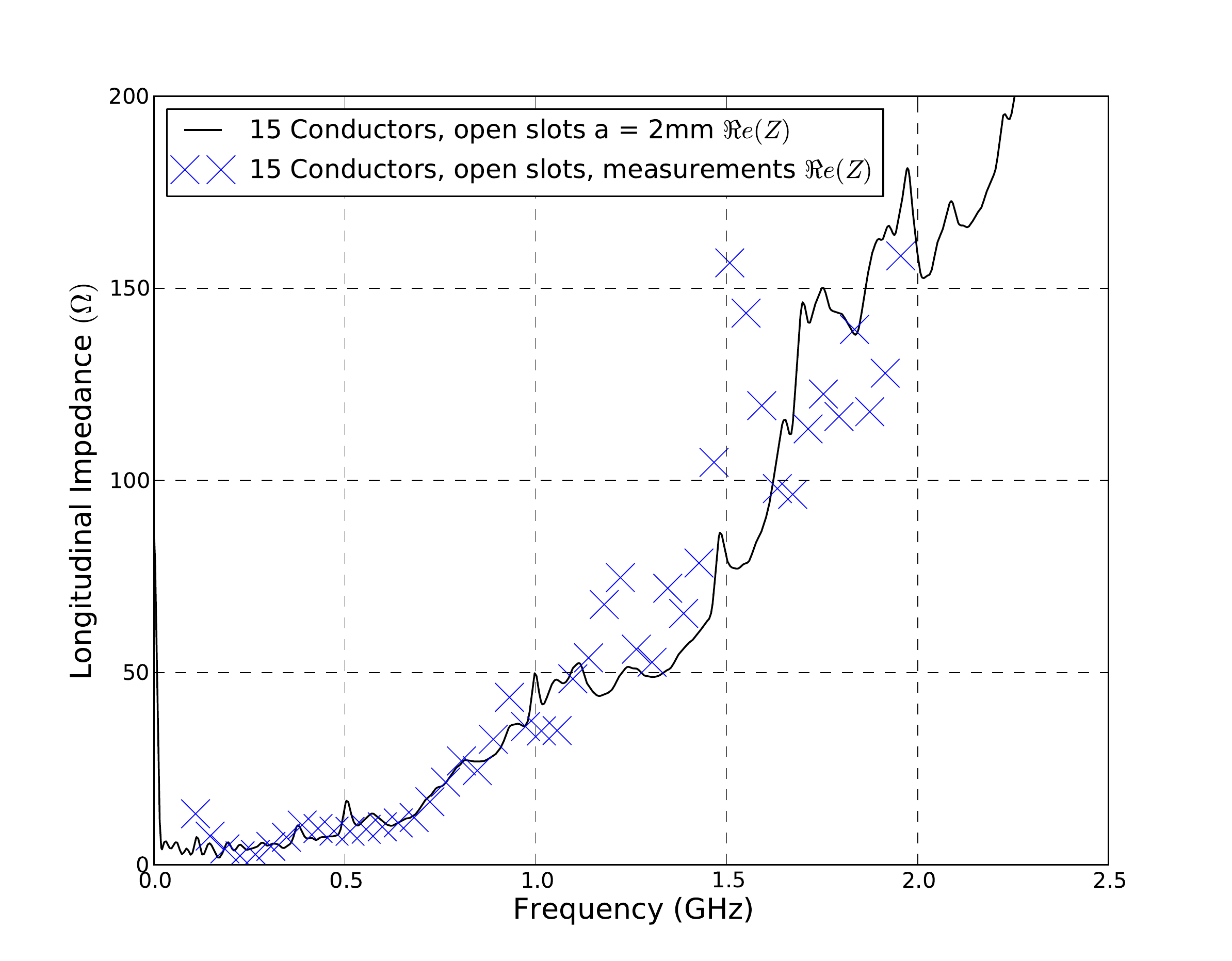}
\end{center}\caption{Comparison between the simulations and measurements of an LHC MKI with a beam screen with 15 screen conductors present in open slots.}
\label{fig:meas_sim_comp}
\end{figure}

To obtain estimates for the beam coupling impedance of the different beam screen designs we have used the time domain simulation code CST Particle Studio\cite{cst-cite}. This code is used to characterise the broadband nature of the expected impedances, which due to their short range wakefields can be efficiently investigated using time domain codes. To ensure the reliability of the complex simulation model, we compare the case of 15 screen conductors to measurements acquired using the stretched coaxial wire technique \cite{kicker_meas}. These results, shown in Fig.~\ref{fig:meas_sim_comp}, show very good agreement between the simulations and measurements below 1.5GHz, with increasing divergence above this frequency due to an incomplete model (RF fingers used to transition from the beam pipe to the beam screen are replaced by a perfect contact) used in the simulations. This level of agreement indicates that the simulation model is suitable to use for reliable and accurate power loss estimates.

\begin{figure}
\begin{center} 
\includegraphics*[width=0.45\textwidth]{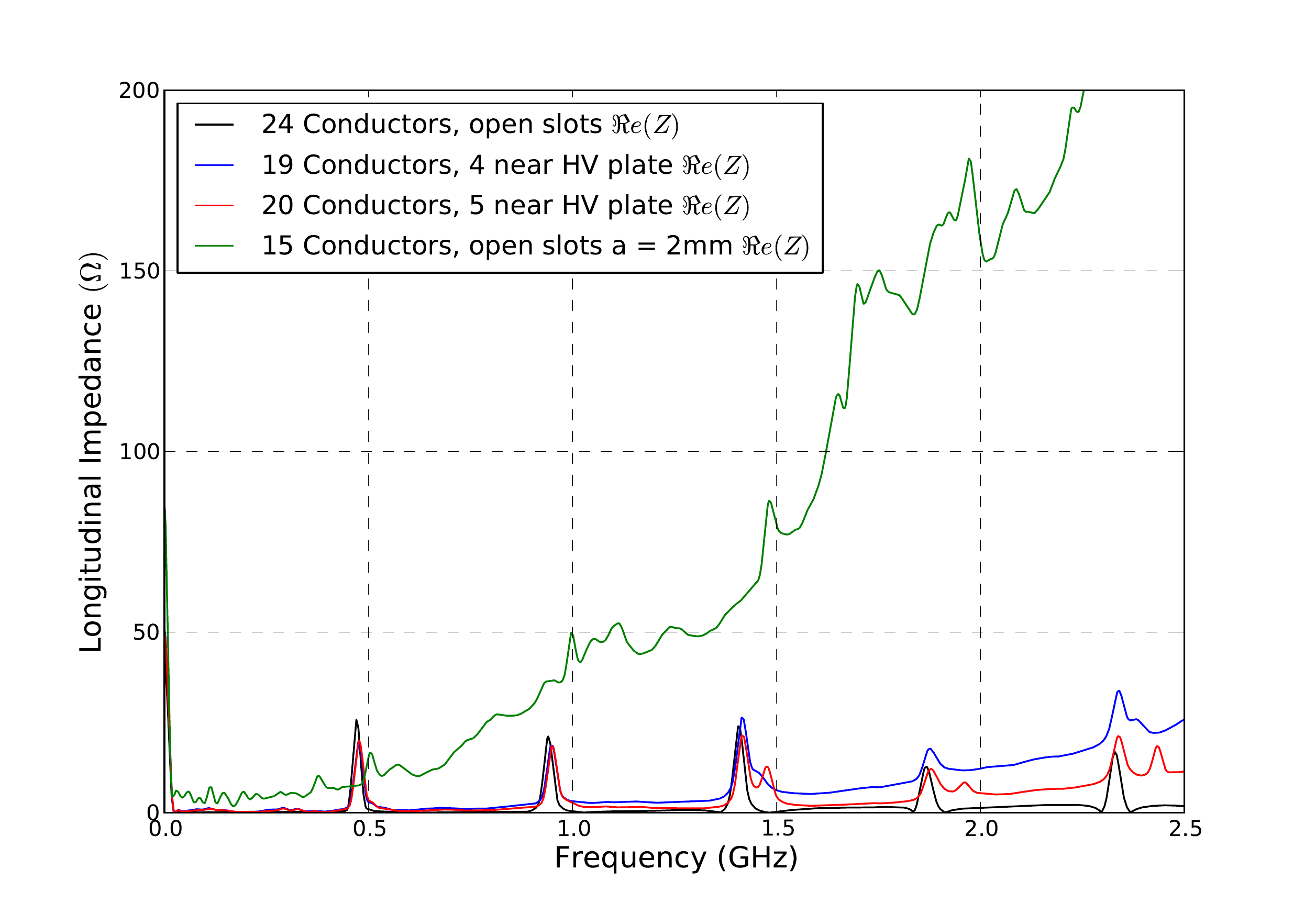}
\end{center}\caption{Simulated longitudinal beam coupling impedance of different numbers of screen conductors in the beam screen of the MKI. Note that even a mild screening of the direction facing the HV plate greatly reduces the beam coupling impedance at low frequencies.}
\label{fig:screen_cond}
\end{figure}

\begin{figure}
\begin{center} 
\includegraphics*[width=0.45\textwidth]{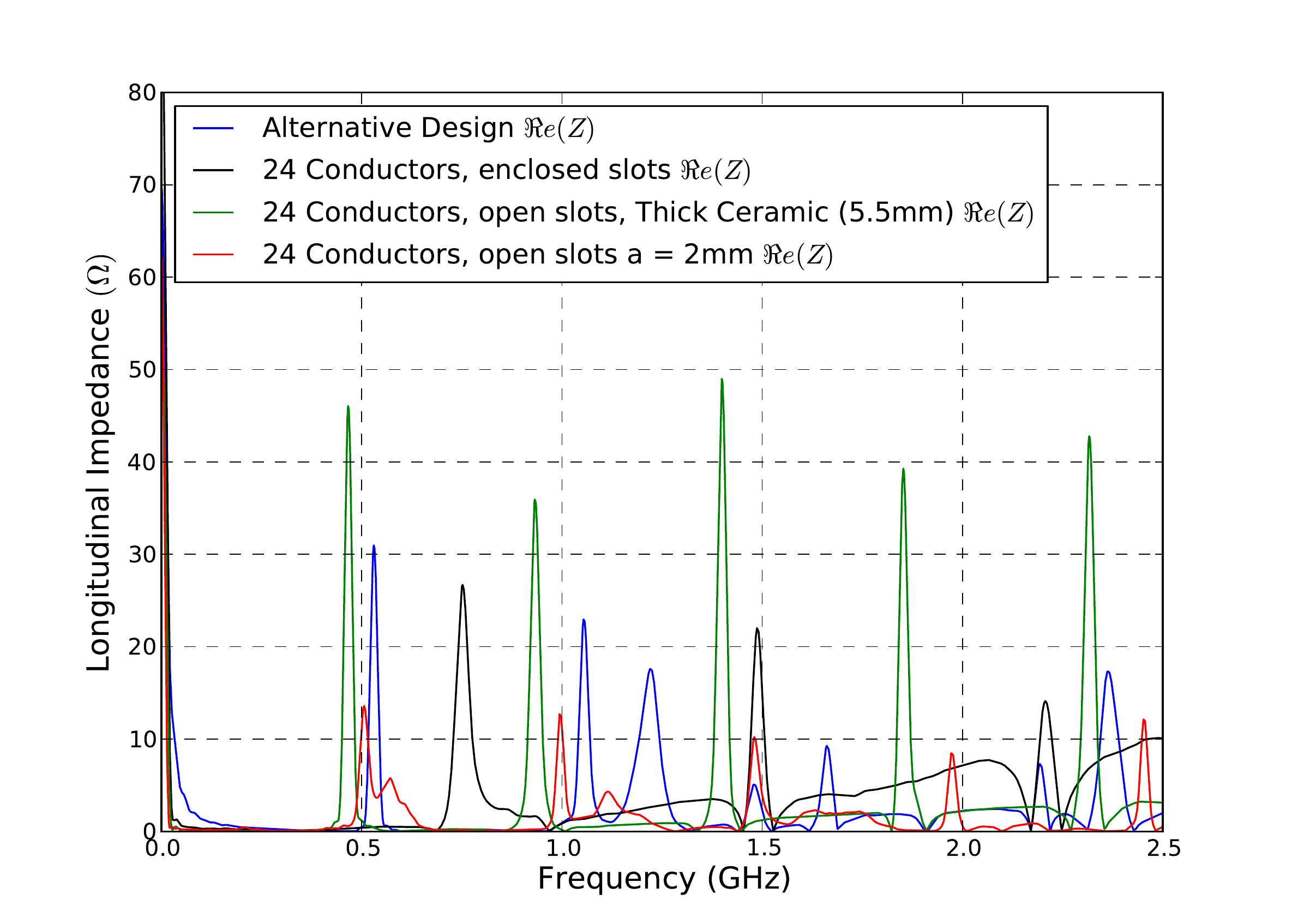}
\end{center}\caption{Simulated longitudinal beam coupling impedance of different screen layouts for the MKI. Note that all have a very similar resonant structure, due to the overlap between the screen conductors and the external metallisation.}
\label{fig:new-screens}
\end{figure}

Using this well understood model, a number of additional screen conductor layouts are investigated using the existing beam screen design: Fig.~\ref{fig:screen_cond} shows the resulting impedance. The additional screen conductors are placed evenly spaced in the area facing the high voltage plate, which can be seen to greatly reduce the impedance at all frequencies compared to having only 15 screen conductors. At low frequencies the impedance tends to the value of having 24 screen conductors. This indicates that at low frequencies even a reduced amount of screeening is very effective at screening the beam from the ferrite yoke, as well as indiciating that the resonant behaviour of the impedance with 24 screen conductors is attributable to the layout of the beam screen.

The longitudinal impedance of alternative screen designs is shown in Fig.~\ref{fig:new-screens} (see Fig.~2 for details). It can be seen that the greatest difference between the different screen designs is the change of the resonant pattern due to the overlap of the screen conductors and the external metalisation on the ceramic screen. This forms a $n\lambda{}/2$ resonator, the characteristic wavelength of which is dependent on the volume of the overlapping section which can change drastically dependent on the screen design.

\section{HEATING ESTIMATES}

\begin{table*}
\begin{center}
\caption{Estimates of the total power loss (Watts) due to the beam coupling impedance of an entire MKI for a number of different beam screen configurations and layouts. Estimates are given with two different bunch lengths. The alternative layout is that shown in Fig.~\ref{fig:alt-screen}. Thick ceramic indicates using a thicker ceramic screen (5.5mm thick as opposed to 4mm for all other designs).}
\begin{tabular*}{0.9\textwidth}{@{\extracolsep{\fill}} c | c | c | c | c }

 & \multicolumn{2}{r|}{Bunch Length $t_{z} = 1.2ns$} & \multicolumn{2}{c}{Bunch Length $t_{z} = 1.3ns$}\\ \hline
Beam Screen Layout & 25ns & 50ns & 25ns & 50ns \\ \hline
No Beam Screen & 4933&2784 &4320 & 2395\\ \hline
24 screen conductors, open &38 &13 &37 & 12\\ \hline
15 screen conductors, open &147 &72 &129 & 60\\ \hline
24 screen conductors, enclosed &76 &23 &73 & 22\\ \hline
24 screen conductors, open, thick ceramic &53 & 20&52 & 19\\ \hline
24 screen conductors, open, alternative layout &63 & 23&62 & 22\\ \hline
19 screen conductors, open &52 &18 &50 & 16\\ \hline
20 screen conductors, open &50 & 16& 48& 15\\
\end{tabular*}
\label{tab:power-loss}
\end{center}
\end{table*}

To ensure a complete evaluation of the impedance profile of the different beam screen configurations we must produce estimates of the beam-induced heating in the LHC-MKIs. This is done by considering the power loss due to the longitudinal impedance of the device.

The power loss $P_{loss}$ due a longitudinal impedance $Z_{\parallel}$ in a storage ring can be given by\cite{metral_cham2012}

\begin{equation}
P_{loss} = \left( f_{rev}eN_{b}n_{bunch}  \right)^{2} \displaystyle\sum\limits_{n = 0}^{\infty} \left( 2 \left| \lambda \left( n \omega_{0} \right)  \right|^{2}  \Re{}e \left( Z_{\parallel} \left(n \omega_{0}\right) \right) \right)
\end{equation}

where $f_{rev}$ is the revolution frequency, $e$ is the electron charge, $N_{b}$ is the bunch population, $n_{bunch}$ the number of bunches in the storage ring, $\lambda\left( \omega \right)$ is the bunch current sprectrum in the frequency domain, $\omega_{0}=2\pi f_{0}$ and $f_{0}~=~\frac{1}{\tau_{b}}$, and $\tau_{b}$ is the bunch spacing.

For heating estimates using the broadband impedance, we can use a variety of possible beam spectra to estimate the power loss. It can be shown that the cos$^{2}$ distribution typically gives a pessimistic estimate of the power loss, and thus is used here to provide conservative estimates of possible heat loads on the kicker magnet. 

For the nominal LHC-type beams, $N_{b}~=~1.15~\times~10^{11}, f_{rev}~\approx~11.8kHz, n_{bunch,25}~=~2808$ for $\tau_{b}~=~25ns$,  $N_{b}~=~1.45~\times~10^{11}, n_{bunch}~=~1380$ for $\tau_{b}~=~50ns$   $f_{0,25}~=~40MHz$ for $\tau_{b}~=~25ns$, $f_{0,50}~=~20MHz$ for $\tau_{b}~=~50ns$.

Table~\ref{tab:power-loss} gives more indepth information for more realistic bunch lengths for 25ns and 50ns schemes.

We can see that those configurations that involve either a bare beam screen or no beam screen at all lead to very large power losses in the MKI, reaching into the kW range for unscreened or badly screened MKIs thus vindicating the choice to implement a beam screen in the LHC-MKI. We see that the implementation of the original beam screen with 24 screen conductors gives a significant reduction in the expected power loss, by a factor of over 100 compared to unscreened ferrite, e.g. 38W for 24 conductors as opposed to 4933W for bare ferrite for 25ns bunch spacing with a bunch length of 1.2ns. The other proposed screen designs give comparable performance to that of the original screen design, giving between 50\% to 100\% additional heating over the original design. In addition to alternative screen layouts, it can be seen that increasing the bunch length can provide a reduction in the expected power loss. Given the assumed bunch spectrum, this reduction is not significant for the layouts showing resonant impedance behaviour ($\approx 4\%$ for 0.1ns), but for the layout with 15 screen conductors the reduction can be significant ($\approx 15\%$ for 0.1ns).

\section{SUMMARY}

A number of beam screen designs have been evaluated to reduce the heat load on the injection kicker magnets for the LHC. The original beam screen design has been shown to result in the lowest expected heat load; but a number of options exist which could produce comparable heat loads, whilst reducing the problem of electrical breakdown between the screen conductors. It has also be shown that it is possible to simulate highly complex structures using modern computational tools. Future work would be to investigate whether the alternative screen designs will contribute negatively to the transverse impedance in the LHC, and also how they affect the rate of electrical breakdown in the MKI.

\end{document}